# A Scalable Resource Management Layer for FPGA SoCs in 6G Radio Units

Nikolaos Bartzoudis, José Rubio Fernández, David López-Bueno, Antonio Román Villarroel
{nikolaos.bartzoudis, pepe.rubio, david.lopez, antonio.roman}@cttc.es
Centre Tecnològic de Telecomunicacions de Catalunya (CTTC-CERCA), Castelldefels, Barcelona, Spain.

*Abstract-* This work presents a perspective on addressing the underutilization of computing resources in FPGA SoC devices deployed in 5G radio and edge computing infrastructure. The initial step in this approach involves developing a resource management layer capable of dynamically migrating and scaling functions within these devices in response to contextual events. This layer serves as the foundation for designing a hierarchical, data-driven micro-orchestrator responsible for managing the lifecycle of functions in FPGA SoC devices. In this paper, the proposed resource management layer is utilized to reconfigure a function based on events identified by a computer vision edge application.

## I. Introduction

System-on-Chip (SoC) devices with field programmable gate array (FPGA) resources are used as function accelerators across the 5G radio access network (RAN) and cloud infrastructures. Taking as a reference the open RAN Alliance (O-RAN) architecture, FPGA SoC devices are encountered i) in open radio units (O-RU) accelerating low physical layer (PHY) digital signal processing (DSP) functions, ii) in network interface cards (NIC) implementing the fronthaul interface, iii) in open distributed units (O-DU) accelerating specific high-PHY functions, and v) in the near-real-time RAN intelligent controller (RIC) hosting the inference of artificial intelligence (AI)/machine learning (ML) models as extended applications (xApp). FPGA SoC devices are also used to accelerate applications across the compute continuum.

The complex architecture of these multi-processing devices (Fig. 1), makes challenging and cumbersome the virtualization of all underlying compute resources, the exposure of deep telemetry data and, consequently, the deployment of such devices in Kubernetes clusters. Equally challenging is the run-time fine grain adaptive management of the computing resources either at task or at function level. Different efforts from the industry and academia have been trying to address these challenges, offering solutions tailored for specific SOC families. For instance, Microsoft's Catapult v2 [1] work focuses on the offloading of network processing from the embedded processor to the FPGA area of the SoC device over Microsoft's Azure framework. Another work in IBM Research divides the FPGA spatially into distinct application regions, where hardware accelerated applications are to be programmed; a service logic secures access to shared off-chip memory and a dedicated server [2]. Amazon's AWS F1 instance offers connectivity to eight FPGA cards which are connected to a single physical server and a dedicated FPGA-only interconnection network [3]. Multiple academic works have also explored the deployment of FPGAs in cloud environments, but their thorough review goes beyond the scope of this paper.

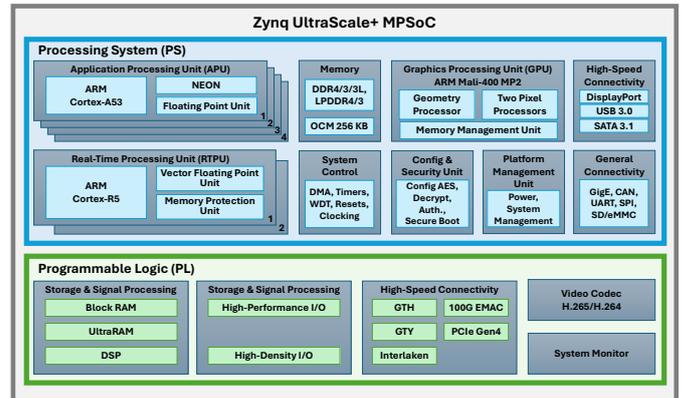

Fig. 1. AMD's Zynq UltraScale+ Multi-Processing SoC (MPSoC) device.

### A. Contributions

This work presents a resource management layer for functions running in FPGA SoC devices, which along with the run-time reconfiguration framework presented in [4] (i.e., joint management of interdependent software and FPGA functions) form the necessary substrate for designing an intelligent closed-loop micro-orchestrator. The goal of the latter is to reconfigure, scale, migrate, or replace functions across the SoC fabric based on different intelligent control loops. Such functionality is facilitated by the resource management layer which allows for the adaptive management of accelerated functions residing in O-RUs and O-DUs, whose rather static operation is expected to be challenged in 6G real-time control loop use cases [5]. The micro-orchestrator could be seen as a hierarchical data-driven intelligent controller that could be built by training an AI/ML model with RAN traffic data, context information and on-chip telemetry data (e.g., execution time, power consumption, throughput of embedded buses). Using the notions of O-RAN, the micro-orchestrator could take the form of a combined rApp and xApp, or a real-time application located at the extreme edge [6].

As part of this roadmap, we present in this paper a FPGA SoC system able to reconfigure its underlying functions based on events detected by a computer vision edge application. This context-driven function reconfiguration is designed with the perspective to be integrated into the low physical-layer (low-PHY) of a commercial O-RU featuring the functional split 7.2.

## II. FPGA Resource Management Layer

The computer vision application is hosted in the AMD Kria KV260 Vision AI Starter Kit [7] and the function that is used for reconfiguration purposes in the AMD Zynq UltraScale+ RFSoC ZCU111 Evaluation Kit [8], thereafter denoted as RU emulation platform. The selected reconfigurable function is a fast Fourier transform (FFT) processing block. This is either hosted in the application processing unit (APU) of the ZCU111 radio frequency SoC (RFSoC) device (i.e., an ARM A53 processor) using the open source FFTW implementation [9], or in the programmable logic (PL) area of the same device using the AMD FFT LogiCore [10]. The computer vision application in the KV260 edge node detects events, which are then communicated to the ZCU111 platform. To do so, a software hook has been added to the KV260 to count and expose events to the ZCU111 platform using a network socket connection.

The FPGA SoC resource management layer is a rule-based controller running under Linux in the APU, able to seamlessly apply: i) function migration, ii) function scaling, iii) function placement, and iv) function reconfiguration. The implementation of the last two options was presented in [4]. As seen in Fig. 2, this work focuses on the run-time function scaling and migration from a software to a hardware-accelerated execution domain, based on events detected by an edge application.

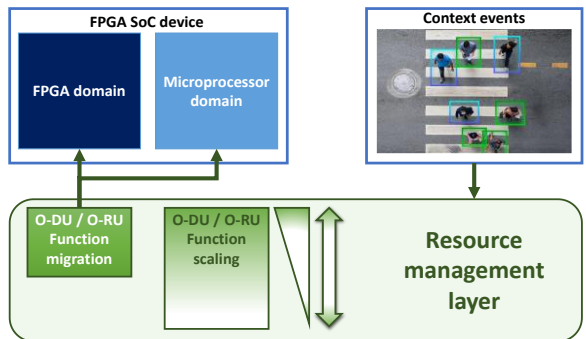

Fig. 2. Function reconfiguration in FPGA SoC devices.

In the following, we describe the hardware, firmware and software components comprising the two main platforms.

### A. Edge node

- PL functions: The PL part of the Zynq UltraScale+ device performs the processing of the video signal. Furthermore, it hosts a natural language processing computer vision application [11], which implements face detection using a deep learning processor (DPU). The latter features the pre-compiled DenseBox face detection model from the Xilinx Vitis-AI Model Zoo (i.e., Network model: cf densebox wider 360 640 1.11G 1.2). The output video signal is overlayed with a frame surrounding each detected face and it is constantly updated in the 2D video space domain.
- Application processing unit (APU) functions: A Linux application hosted in the APU configures and initializes the PL part. On top of that, a module was added to process the DPU output and count the number of detected faces. Upon event occurrence, the number of detected faces is notified to the ZCU111 RFSoC device. To do so, an APU application sends a message through a socket network connection between the KV260 and ZCU111 boards.

### B. RU emulation platform

- PL functions: The PL accelerated FFT [10] uses an input signal playbacked from the platform's DDR memory. The output of the FFT is stored in another area of the DDR memory. The FFT processing block is always configured in the PL area, but when not in use, it is deactivated through clock-gating signalling, to reduce the PL dynamic power consumption.
- APU functions: A socket client application receives the messages from the KV260 board and retrieves the issued events (number of faces detected). The APU executable includes a precompiled FFTW. Switching between this FFT software version and the FPGA-accelerated PL FFT version is made feasible by a reconfiguration controller (i.e., part of the FPGA SoC resource management layer), which performs FFT function migration plus scaling at run-time (i.e., variation of the number of points of the FFT). The reconfiguration controller takes the following actions upon event detection: i) if 0 faces are detected then the FFTW is used (8 points FFT), ii) if 1 face is detected the FFTW is used (FFT scales to 1024 points), iii) if 2 faces are detected the FFT LogiCore is used in PL (and FFT scales to 2048 points), iv) if more than 2 faces are detected the FFT LogiCore is used in PL (and FFT scales to 4096 points). The execution time of the FFT at the PL and APU is calculated and shown on-screen (Fig. 3, left) A performance comparison is also applied by calculating the mean squared error between the software-executed FFTW function (floating point operations) and the FPGA accelerated FFT function (fixed-point precision).
- Power monitoring: An application was created to monitor the power consumption of the ZCU111 RFSoC device. The embedded Linux system running in the APU periodically reads the monitoring data from the on-chip voltage sensors and dumps into the Linux file system. The power monitoring application sends the metrics to a host machine through a socket network connection, where they are visualized in a Python application (Fig. 3, right). As it is observed, different voltage rails of the RFSoC device corresponding to different processing elements and peripherals are monitored.

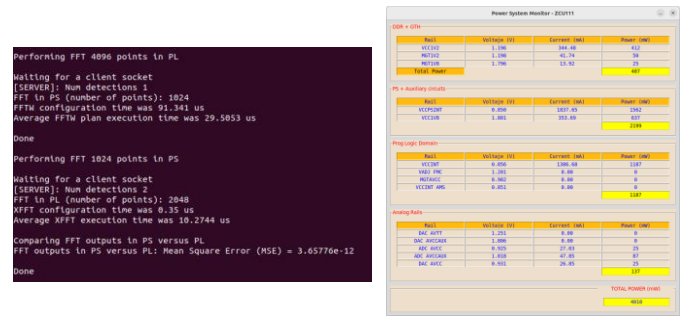

Fig. 3. Metrics visualization from the experimental setup.

The complete experimental setup is shown in Fig. 4.

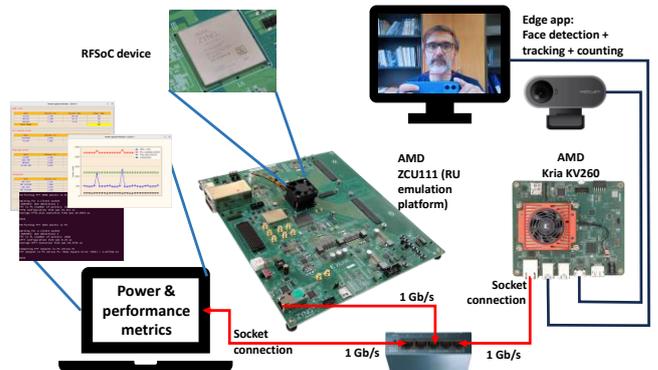

Fig. 4. Hardware diagram of the experimental setup.

## III. RESULTS AND DISCUSSION

A time measuring utility was engaged to calculate the execution time in both the APU and PL implementation of the FFT function, placing markers and counters at the input and output of the PL-implemented FFT function and at the software buffer that processes the input data in the case of the APU-based FFTW function. Table 1 summarizes the execution time difference of the PL-based FFT and the FFTW for the four sizes of the FFT function we considered in our experimental setup analysis. In brackets the values not evaluated in the real-time system but extracted through dedicated measurements. The execution time in the PL is deterministic, whereas in the case of the APU, the execution time variates due to the execution variability of the embedded operating system (i.e., a Linux distribution of AMD for RFSoC devices called Petalinux); thus, the average of 20 measurements has been applied to the APU results.

Table 1. Comparison of APU versus PL FFT function execution time.

| FFT points | APU execution time (us) | PL execution time (us) | Acceleration factor |
|---|---|---|---|
| 8 | 0.28 | (0.04) | 7.9 |
| 1024 | 50.62 | (5.45) | 9.3 |
| 2048 | (113.55) | 8.7 | 13.1 |
| 4096 | (278.72) | 18.07 | 15.4 |

A similar testing was carried out to measure the total power consumption of the FPGA SoC device. To do so, the AMD's FPGA SoC power management framework was leveraged; the latter provides a centralized power state information through the use of a platform management unit (PMU) and supports energy management APIs. The power management framework also provides support for Linux power management (e.g., including Linux device tree power management, power management states), along with direct power management control for memories and peripherals. For the concrete needs of this this integration and validation stage, the SDRAM, APU and PL voltage rails along with the current were monitored. It is important to note that the data is retrieved by an external SDRAM memory and the execution of the FFT lasts for a predefined time length (i.e., equal to the amount of data retrieved by the SDRAM). As expected, the use of the SDRAM during the offloading process from the APU to the PL and vice versa impacts its power footprint. Table 2 provides a summary of the power consumption results per processing element when the FFT function is scaled and migrated from the APU to the PL; in brackets the static power consumption of the PL and APU when not hosting the FFT function.

Table 2. Breakdown of SDRAM, APU and PL power consumption, upon FFT offloading based on context events.

| | Power consumption (mW) | | | |
|---|---|---|---|---|
| FFT points | DDR | APU | PL | Total |
| 8 | 425 | 2064 | (1187) | 3676 |
| 1024 | 537 | 2224 | (1187) | 3948 |
| 2048 | 965 | (2024) | 1365 | 4354 |
| 4096 | 1358 | (2024) | 1584 | 4966 |

### A. Applicability of iFFT offloading to 5G NR low-PHY

The final objective of the context driven FPGA SoC function offloading presented in this paper is to be integrated in the low-PHY layer of a 5G NR Radio Unit (RU). A crucial aspect to consider in such a real-life scenario is that the FFT function offloading in a FPGA SoC device is bound to the maximum FFT size that can be processed in the APU without violating the required time to process the I and Q samples in a given 5G NR slot. Thus, taking as example a 30 *kHz* subcarrier spacing configuration in 5G NR, the goal is to be able to process 20 *MHz* signals that require a 1024-points FFT [12] (Table B.5.2-2). In such a real-time system the typical partitioning of functions between the PL and the APU is the following:

- The PL area can efficiently handle tasks like OFDM modulation/demodulation, cyclic prefix (CP) addition/removal, channel estimation and equalization, resource mapping and demapping, precoding and beamforming.
- The APU must complete the FFT processing within the real-time constraints of the 5G NR slot structure. A 1024-point FFT requires moderate computational effort, especially when optimized (e.g., FFTW function or custom NEON-optimized FFT libraries - NEON is a vector co-processing extension of the APU-).

An efficient function offloading also depends on the data transfer from the PL to the APU and vice versa. In the following, the key required steps that enable such data transfers are presented.

**Processing flow**

The PL processes incoming data (e.g., resource mapping, channel modulation) to generate frequency-domain symbols for the inverse FFT (iFFT). These symbols are transferred to the APU via an AXI4-Stream interface, using an AXI DMA (Direct Memory Access) controller to bridge streaming data to memory-mapped transactions. The DMA writes the data to a physical address range in the on-chip memory (OCM) of the FPGA SoC (RFSoC), where the APU accesses it through memory-mapped reads over the AXI interconnect. Depending on requirements, the APU can use either the accelerator coherency port (ACP) for cache-coherent access or an AXI high-performance (HP) port for lower-latency transfers. Memory pooling in OCM is critical to prevent overwrites and minimize latency.

After the APU computes the iFFT, it writes the resulting time-domain samples back to a shared OCM buffer via its AXI Master interface. The PL then retrieves this data using an address-aware AXI Master interface to proceed with low-PHY operations (e.g., cyclic prefix addition, DAC interfacing). Synchronization between PL and APU is handled through interrupts triggered by DMA completion flags—for instance, the PL interrupts the APU when frequency-domain symbols are ready, and the APU signals the PL after processing.

This end-to-end offloading process requires careful design to meet timing constraints, leveraging AXI DMA for efficient transfers and OCM-optimized buffering to reduce overhead. A high-level overview of the PL-APU interfacing to enable the iFFT offloading is shown in Fig. 5.

**Latency budget analysis**

An analysis of the latency budget requirements for a 20 *MHz* (30 *kHz* SCS) 5G NR signal is explained in the following:
- Slot duration: 0.5 *ms* (14 OFDM symbols/slot).
- Symbol duration: 35.7 *μs* (CP included).
- iFFT deadline: Within less than 35.7 *μs* (ideally less than 20 *μs* to allow for other low-PHY processing).

The offloading of the iFFT from the PL to the APU in AMD MPSoC/RFSoC devices is feasible for 5G NR systems with a

20 *MHz* bandwidth and 30 *kHz* subcarrier spacing, provided strict latency constraints are met. The 1024-point iFFT must be completed within the 5G NR symbol duration of 35.7 *μs*, which implies optimized data transfers and computation tasks. The PL processes frequency-domain symbols (e.g., after resource mapping and modulation) and transfers them to the OCM via an AXI DMA controller configured for Scatter-Gather mode. Theoretical and empirical measurements indicate that transferring 8 KB of data (i.e., 1024 complex single-precision floats) over a 64-bit AXI HP port at 300 *MHz* achieves an effective throughput of approximately 1.6 *GB/s*, resulting in a transfer latency of approximately 5 *μs* [13] (i.e., 8KB: 8192 bytes / $1.6 \times 10^9$ *bytes/s* = 5.12 *μs*). This aligns with benchmarks that report real-world DMA throughput tests under non-cacheable burst conditions [14]

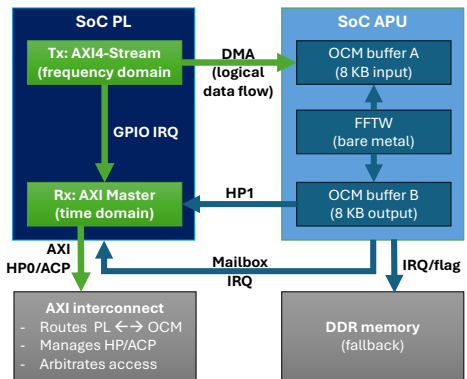

Fig. 5. PL-APU interfacing to enable iFFT offloading.

The APU then executes the iFFT using an FFTW3 library optimized for ARM NEON instructions. In bare-metal environments where the OS jitter and additional OS-related top-up latencies do not apply, a 1024-point complex FFT completes in less than 10 *μs* on ARM Cortex-A53 cores [15]. This performance assumes precomputed twiddle factors and cache locking to prevent thrashing, as outlined in the Xilinx APU Optimization Guide [16] (i.e., the precomputed Plans FFTW_MEASURE reduces runtime latency by approximately 30%). The resulting time-domain samples are written back to OCM, which the PL reads via AXI HP with a similar 5 *μs* latency, leveraging zero-wait-state OCM access [17].

Synchronization between PL and APU is achieved through GPIO PL-to-APU interrupts and Mailbox-driven APU-to-PL signalling. Bare-metal implementations reduce interrupt latency to less than 1 *μs* [17]. The total worst-case latency of 21 *μs* leaves a 14.7 *μs* margin for CP insertion and DAC interfacing, ensuring compliance with 5G NR timing requirements [18]. The estimation of the key latency budgets is summarized in the Table 4.3 that follows.

Table 3. Key latency budget for the FFT offloading.

| Steps | Latency target (in μs) |
|---|---|
| PL → OCM (Freq Symbols) | ≤ 5 |
| APU iFFT (FFTW) | ≤ 10 |
| OCM → PL (Time Samples) | ≤ 5 |
| Interrupts | ≤ 1 |
| Total iFFT Offload | **≤ 21** |

The validation of this design requires profiling DMA transfers using the AXI performance monitor (APM) [19], measure the FFTW execution time with cycle-accurate timers and verify OCM access patterns using memory debug tools to confirm absence of contention.

*B. Other considerations*

Integrating the FPGA SoC reconfiguration framework [4] in the resource management layer is also bound to some limitations that need to be carefully considered. According to the FPGA SoC boot time estimation tool [20] a partial PL reconfiguration bitstream occupying for instance 5% of the ZCU111 flash memory, would approximately require 10 *ms* to be transferred in a PL reconfigurable region. Thus, faster reconfiguration strategies like the activation/deactivation of PL functions, or longer-term FPGA SoC resource usage forecasting shall be contemplated in real-life use cases. A similar procedure is followed for the APU-to-PL offloading.

## IV. CONCLUSIONS

This work lays the foundation for more sustainable use of processing elements in FPGA SoC devices at the far and extreme edge, where computing resources are limited, and energy and latency constraints are critical. Function offloading enables multitenancy, optimizing resource utilization. Offloading the FFT processing to the APU frees up PL resources for other edge applications enabling new 6G network use cases. The proposed framework could be also extended to high-PHY functions like channel coding and enhance FPGA SoC resource management by incorporating the ARM R5 processor and the ARM A53 NEON instructions.

ACKNOWLEDGEMENT

*The work in this paper was supported by the project 6G-RASING (PID2023-146245OB-C22) funded by the Agencia Estatal de Investigación (MCIN).*